# Magnetic Frustration in a Zeolite

Danrui Ni[a], Zhiwei Hu[b], Guangming Cheng[c], Xin Gui[a], Wenzhu Yu[d], Chunjiang Jia[d], Xiao Wang[b], Javier Herrero-Martín[e], Nan Yao[c], Liu Hao Tjeng[b], and Robert J. Cava*[a]

[a]Department of Chemistry, Princeton University, Princeton, NJ 08544, United States

[b]Max Planck Institute for Chemical Physics of Solids, 01187 Dresden, Germany

[c]Princeton Institute for the Science and Technology of Materials, Princeton University, Princeton, NJ 08544, United States

[d]Key Laboratory for Colloid and Interface Chemistry, Key Laboratory of Special Aggregated Materials, School of Chemistry and Chemical Engineering, Shandong University, Jinan 250100, PR China

[e]ALBA Synchrotron Light Source, E-08290 Cerdanyola del Vallès, Barcelona, Spain

*Corresponding author. E-mail address: rcava@princeton.edu

## Abstract

Zeolites are so well known in real world applications and after decades of scientific study that they hardly need any introduction: their importance in chemistry cannot be overemphasized. Here we add to the remarkable properties that they display by reporting our discovery that the simplest zeolite, sodalite, when doped with $Cr^{3+}$ in the β-cage, is a frustrated magnet. Soft x-ray absorption spectroscopy and magnetic measurements reveal that the Cr present is Cr(III). Cr(III), with its isotropic $3d^3$ valence electron configuration, is well-known as the basis for many geometrically frustrated magnets, but it is especially surprising that a material like the $Ca_8Al_{12}Cr_2O_{29}$ zeolite is a frustrated magnet. This finding illustrates the value of exploring the properties of even well-known materials families.

Zeolites are a very important family of crystalline microporous solids, described even in rudimentary chemistry texts.[1] They have been widely studied from the chemical, geometrical, topological, and crystallographic perspectives. Among the multi-faceted studies of the chemical, structural, physical, and mechanical properties of zeolite-based materials, the behavior of these cage-structured compounds when magnetic centers are present (e.g. transition metals with unpaired electrons, or alkali-metal ionic clusters that trap single electrons) has also attracted attention, in spite of the fact that it is the framework structures of zeolites are the characteristic that makes them the most useful.[2–4]

Sodalite Zeolites, materials with the general formula $\boldsymbol{M}_8(\boldsymbol{T}_2O_4)_6\boldsymbol{X}_2$ (where $\boldsymbol{M}$ = alkali/alkali earth metals, transition metals, rare earth metals, or some organic cation groups; $\boldsymbol{T}$ = Si, Al (and/or other species that make up the framework of corner sharing tetrahedra; $\boldsymbol{X}$ = mono- or di-valent anions), are the simplest members in the enormous zeolite family, having the simplest charge-balanced crystal structure[1,5] The "sodalite cage" or "β-cage", surrounded by the cuboctahedron framework, forms the basis of many zeolites when stacked with other framework geometries to create different arrangements of cages or tunnels. In sodalite itself, the simplest zeolite, the joining of the β cages yields a body centered cubic lattice with hexagonal apertures between cavities. Studies have been carried out to investigate the electronic and magnetic properties of sodalites, leading to the discovery of some new ferromagnetic or antiferromagnetic sodalite-type phases.[6–9]

The Calcium aluminate chromate sodalite $Ca_8Al_{12}O_{24}(CrO_4)_2$ (also written as CACr), is a typical sodalite compound, with $\boldsymbol{M}$ = Ca, $\boldsymbol{T}$ = Al, and $\boldsymbol{X}$ = $CrO_4^{2-}$. It is known as a "Si-free sodalite" and has been extensively studied for its structural properties, including its phase transitions, structural modulations and twining, by neutron diffraction, in-situ synchrotron diffraction, and selected area electron diffraction.[10–14] The structure of its *I*-43*m* unit cell is shown in Figure 1A (This is the traditional view, with Al is at the vertices where the line segments of the framework meet, and the oxygens, which are at the midpoints of the line segments, omitted from the drawing. Both the Ca and the $CrO_4^{2-}$ groups are inside the β-cages, as described further below). This is considered as the high-temperature phase of CACr, as the compound has been reported to undergo weakly distortive phase transitions through orthorhombic, tetragonal, and pseudo-cubic symmetries at lower temperatures evidenced primarily by weak superstructure reflections. The orthorhombic and tetragonal superstructures have not been quantitatively determined, although they have been observed in the electron microscope.[13,15] The complex superstructures observed have been attributed to orientational ordering of the $CrO_4^{2-}$ anions in the cages. The anions are located in the centers of the β-cages and are coordinated to four $Ca^{2+}$ cations that sit near the centers of the hexagonal faces of the cages. The chromium in the $CrO_4^{2-}$ group is in a +6 chemical valence state and therefore has no unpaired electrons. The compound is thus expected to be diamagnetic.

Sulfate- and selenite-bearing sodalites can be reduced under hydrogen, generating sulfide- and selenide-containing sodalites without breaking up the framework. Therefore, we considered it feasible to maintain the sodalite framework while reducing the Cr(VI) in the cages to Cr(III) and to thereby produce magnetic centers in the structure. If the chromium in the system could be reduced from Cr(VI) to Cr(III), then unpaired electrons will be introduced, which has the potential to result in interesting magnetic behavior due to the distinctive β-cage framework geometry, the possible orientational order or disorder of Cr(III)$O_6$ groups in the cages, and the electron configuration of Cr(III). Cr(III) with its $d^3$ electron configuration is a well-known isotropic spin ion when in an octahedral crystal field; its most commonly known coordination environment in oxides.

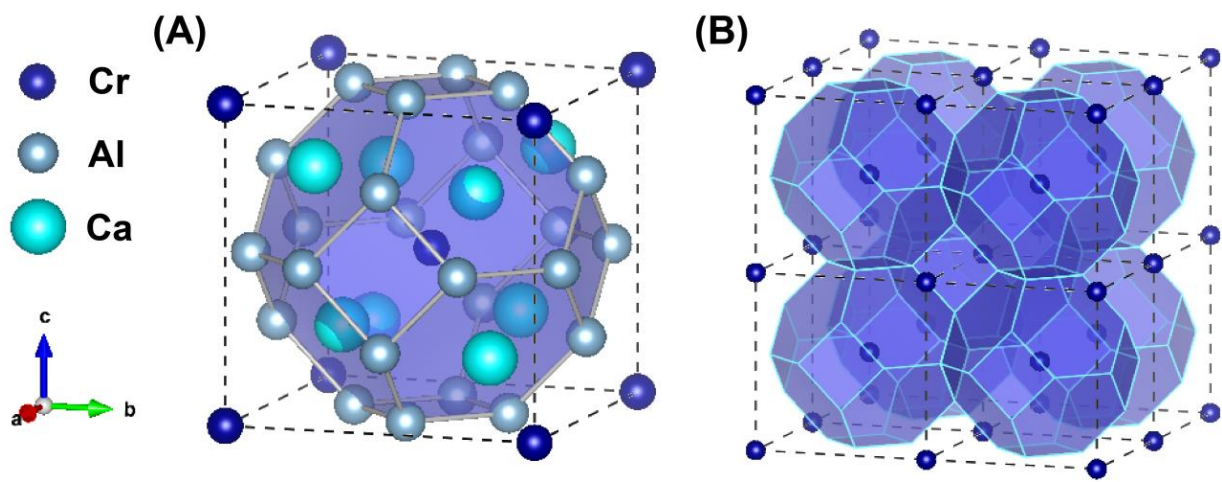

**Figure 1.** The *I*-43m unit cell of the CACr sodalite, emphasizing the sodalite β-cage, which is shaded in blue. Oxygens are omitted from the representation which shows only the Al positions on the sodalite cage, the Ca in the hexagonal rings of the beta cages and the Cr's in the ideal BCC arrangement in the cavities (A) A single unit cell (B) 2 x 2 x 2 unit cells revealing the connections of β-cages, with all atoms omitted from the figure with the exception of Cr.

The stoichiometric CACr sodalite was initially prepared by annealing $CaCO_3$, $Al_2O_3$ and $Cr_2O_3$ in air at 1100 °C, with intermediate grindings until the reaction was complete. The polycrystalline samples of the CACr sodalite are bright yellow, and its powder X-ray diffraction (XRD) pattern is in good agreement with the diffraction pattern in the cubic I-43*m* phase of CACr (Figure 2A) in the powder diffraction database.[13] The tiny peaks at low angle may be attributed to a superstructure from the ordering of the $CrO_4^{2-}$ anions in the cages. The as-made CACr sodalite was then annealed at 750 °C under hydrogen gas flow for two days. The resulting powders showed a dramatic color change - from bright yellow to bright green. The bright green color is characteristic of Cr(III) in oxides in highly concentrated form. This obvious color change can be realized under reducing gas flow at an annealing temperature as low as 600 °C, which is lower than the formation temperatures of other possible calcium aluminate or calcium chromate impurities. Several new peaks showed up in the low angle part of the X-ray powder diffraction pattern of the reduced phase (labeled with stars in the blue pattern of Figure 2A). Given that the *d* values of these peaks are related to those of the as-made CACr sodalite, we infer that they were formerly forbidden by the symmetry of as-made sodalite; symmetry that is changed by the removal of the oxygen and the creation of favorable Cr(III)-O bonding environment in the cages. The reduction process has little effect on the structure of the framework, and thus the basic framework diffraction peaks do not shift their positions significantly. The powder XRD analysis therefore confirms that intracage reduction breaks the original symmetry and does not change the cage framework size.

The reduced Cr-sodalite samples were examined by transmission electron microscopy. Electron diffraction analysis (Figure 2B) reveals that only sodalite derived-phases exist in the reduced Cr-sodalite samples. No impurity phases are detected. Two of the sodalite-derived phases are simply superstructures of the CACr sodalite (Phase I at Figure 2B top image with *a* = 9.22 Å, and Phase II in Figure 2B middle with *a* = 13.04 Å = $\sqrt{2}$ x 9.22 Å), suggesting that they are primarily due to different ordering schemes of the Cr-O polyhedra in the cavities. The third sodalite-derived phase is body-centered cubic (*a* = 9.22 Å, Figure 2B bottom), which corresponds to the cell of the as-made yellow Cr-sodalite. Phase II is metastable under the high energy electron beam and easily transforms to a more stable Phase I, before all phases change to the *a* = 9.22 Å body-centered cubic phase after significant exposure to the electron beam. The orientational ordering/disordering transition of Cr-centered polyhedra in the sodalite cage, to which we attribute the superstructures of the reduced CaCr sodalite, has also been observed and reported in the Cr(VI) yellow CACr sodalite.[10,13] According to the powder XRD patterns of the reduced samples, as shown in Figure 2A (blue curve) the primitive cubic phase sodalites are dominant in the reduced Cr-sodalite. The uniform distribution of the metal elements in the different sodalite phases are confirmed by STEM-EDS element mapping and high resolution TEM, as shown in Figures 2C and S1

Borothermal and carbothermal reduction yielded the same results as heating in $H_2$, indicating that the change in Cr oxidation state is due to the change in oxygen content and not due to the exchange of oxygen with $H^-$ (Figure S2). The reduced material returns to a yellow color and its original diffraction pattern after heating in flowing $O_2$ at 900 °C for 2 days. A two-day anneal under the low temperature of 750 °C under an oxygen flow only turns the reduced green powder to an olive color instead of the bright yellow one. These results reveal that the re-oxidation process, which requires in-diffusion of oxygen into the cavities of the reduced material and its ionization, is more difficult than the reduction process. In other words, that it is more difficult for oxygen to be inserted into the cage than to be removed.

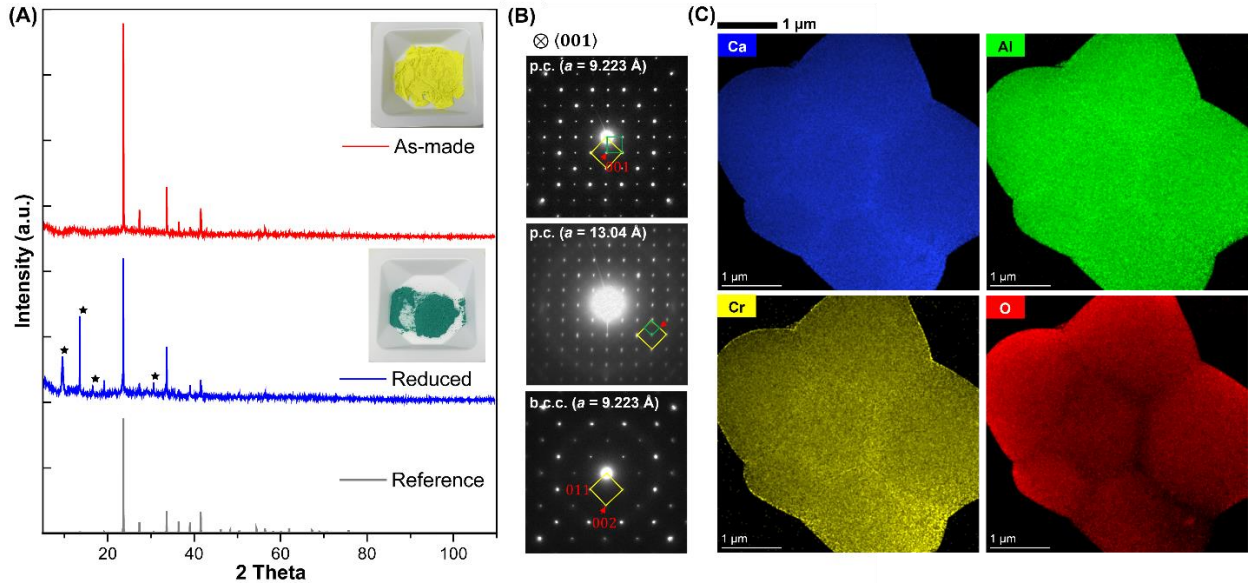

**Figure 2.** (A) Powder XRD patterns of the as-made and reduced Cr-sodalite samples, with the images in the insets showing the color change (B) The electron diffraction patterns of the reduced Cr-sodalite samples along the <001> direction, revealing the primitive cubic (p.c.) superlattice phases Phase I (top), Phase II (middle), and the body-centered cubic (b.c.c.) phase on the bottom; (C) STEM-EDS elemental maps of reduced Cr-sodalite particles.

The temperature-dependent magnetic susceptibility ($\chi = M/H$), measured from 1.8 K to 300 K under an applied field of 5 kOe for both the as-made CACr sodalite and the reduced green sodalite, is shown in Figure 3A. This data reveals the presence of the magnetic frustration in the reduced material. The as-made CACr sodalite, as expected, shows diamagnetic behavior over nearly the whole temperature range of the measurement, except for the appearance of a very small paramagnetic tail below 3 K due to an extremely small number of impurity spins in the sample. This is entirely consistent with the fact that the Cr(VI) in the $CrO_4^{2-}$ ion is in the 6+ state and therefore has no unpaired $3d$ electrons present. In contrast, strongly paramagnetic behavior is recorded for the reduced green sodalite, consistent with the introduction of unpaired electrons. These unpaired electrons are easily attributed to Cr(III) as it is the only transition element present. Fitting of Curie-Weiss law to the data in the range of 175-297 K results in a $C$ of 2.19 (emu Oe$^{-1}$ mol$_{Cr}^{-1}$ K$^{-1}$) and $\vartheta$ of -255 K, with $\chi_0$ = -0.0015 emu/mol$_{Cr}$. The large negative Curie-Weiss $\vartheta$ suggests the dominance of strong antiferromagnetic coupling between the spins in the Cr(III) sodalite in the temperature range of 175–300 K. To further demonstrate the magnetic behavior, $1/(\chi - \chi_0)$ is plotted versus $T$ in the inset of Figure 3A, revealing linear behavior in the high temperature range, as described above, but a continuous change in slope at lower temperatures. This change in slope can either be due to a spin state transition of the Cr(III) or a change in the balance between antiferromagnetic and ferromagnetic interactions between the spins in their fluctuating state, i.e. magnetic frustration. A Curie-Weiss $\vartheta$ of -1.4 K, two orders of magnitude smaller than is seen at higher temperatures, is found by fitting the data in the temperature range 2-10 K. The deviation at low temperature is reminiscent of a presence of increasing ferromagnetic fluctuations with decreasing temperature, and magnetic frustration, an observation that is surprising considering how frequently the sodalites have been previously studied. No magnetic ordering or spin freezing is observed for the green Cr(III)-based reduced CACr down to 1.8 K under 5 kOe, in spite of the very large antiferromagnetic Curie-Weiss $\vartheta$ of -255 K determined from the high temperature fit.

The $1/(\chi - \chi_0)$ curve allows for one to determine the effective magnetic moment of the magnetic ion present. Thus $\mu_{eff}$ is calculated to be 4.19 $\mu_B$ per Cr, which, compared to 3.87 $\mu_B$ for $S$ = 3/2, is in conformity with the expectation for the fact that Cr(III) is the magnetic species present in the reduced sample. A plot of $\chi T$ versus $T$ of the reduced Cr-containing sodalite, presented in Figure 3B, shows a continuous change with temperature, and a trend to saturation in the high $T$ range, confirming the reliability of the effective moment obtained from high temperature data fitting [16].

The field-dependent magnetization was measured on varying the external field from -9 T to 9 T at 2 K and 200 K (Figure 3B inset). It shows a linear dependence at 200 K, and an S shape saturating at around 0.58 $\mu_B$/mol$_{Cr}$ under a magnetic field of 9 T at 2 K. This saturation is likely a representation of how the ferromagnetic components of the Cr(III) spins freeze in place at high applied fields at low temperatures.

There are some cases where frustrated magnetism is not due to the geometric arrangement of the spins, but also can come from frustrated spin configurations and other types of fluctuations[17]. One of the possible explanations for the frustrated magnetic behavior observed in our reduced sodalite samples is disorder of the reduced Cr(III)-centered polyhedra in the sodalite cages. With the possibility of multiple different orientations of the Cr-O polyhedra, the magnetic ions may constitute a disordered magnetic system that "frustrates" the antiparallelism of adjacent spins at low temperatures, resulting in the inability of the system to find a single lowest energy ground state. "Spin glass"-like behavior results in such cases– as in a spin glass, no single configuration of the spins is uniquely favored due to the disorder, which results in coexisting interactions of different strengths, and magnetic moments that are frozen in random directions in the lowest energy state[18]. This may be the case in the reduced Cr zeolite. At temperatures above 1.8 K there is no sign of a frequency dependent magnetic susceptibility, the classical indication of the presence of a spin-glass state, evidence for the eventual freezing of some of the spin components is found in the lower temperature heat capacity characterization, described in the following.

With a rewritten Curie-Weiss Law (Equation (1)), the magnetic behavior for different substances can be compared. This is straightforwardly seen by normalizing the absolute temperature by the Curie Weiss theta (i.e. T/|$\vartheta$|) and also by the strength of the magnetic interactions (|$\vartheta$|) and the magnitudes of the moments ($C$). The result is shown in figure 3D, which is very strong evidence

for the magnetic frustration in the Cr-zeolite. The *C* and θ values used in the plot are those obtained from the fitting of Curie-Weiss Law to the high-temperature data[19]. This type of plot is particularly useful for studying frustrated magnetism, as the *x*-axis ($T/|\vartheta|$) corresponds to the inverse of the frustration index *f*.[20] The thus normalized inverse susceptibility, and the normalized low-temperature heat capacity are plotted in Figure 3D, for reduced CACr sodalite as well as some representative chromium-containing compounds with well-established geometrically frustrating lattices. These spinel and ferrite-structure chromates are classic examples of frustrated Cr oxides and have been studied for decades[21–23]. Ideal antiferromagnetic Curie-Weiss behavior is labeled as a grey straight line in the figure. All the frustrated chromates and our magnetic zeolite are concurrent at temperatures higher than the curie Weiss theta, as expected; the action begins for temperatures below the Curie Weiss theta. With the exception of the chromite spinel, the ferrite-based chromates show negative deviations from the grey line, suggesting significant ferromagnetic deviations from the Curie-Weiss law at low temperatures, with an increasing magnitude of deviation from the simple antiferromagnetism-dominant Curie Weiss behavior at low temperatures compared to the Curie Weiss theta appearing to be largest for the zeolite.

$$\frac{C}{|\theta|(\chi-\chi_0)} = \frac{T}{|\theta|} + 1 \qquad (1)$$

Finally, a temperature-normalized heat capacity comparison is shown in the inset. In this type of plot, the area under curves is a measure of the magnetic entropy for each of the comparison compounds. Two features are immediately apparent, one being that the systems lose magnetic entropy at very low temperatures, less than $T/|\vartheta|$ = 0.1, and that the zeolite loses its magnetic entropy at the lowest relative temperature of all, at about 0.02 $T/|\vartheta|$). Second Is that the zeolite loses its magnetic entropy over a narrower range of temperature than the chromate ferrites, with the exception of $ZnCr_2O_4$, where a structural phase transition accompanies the long range magnetic ordering.

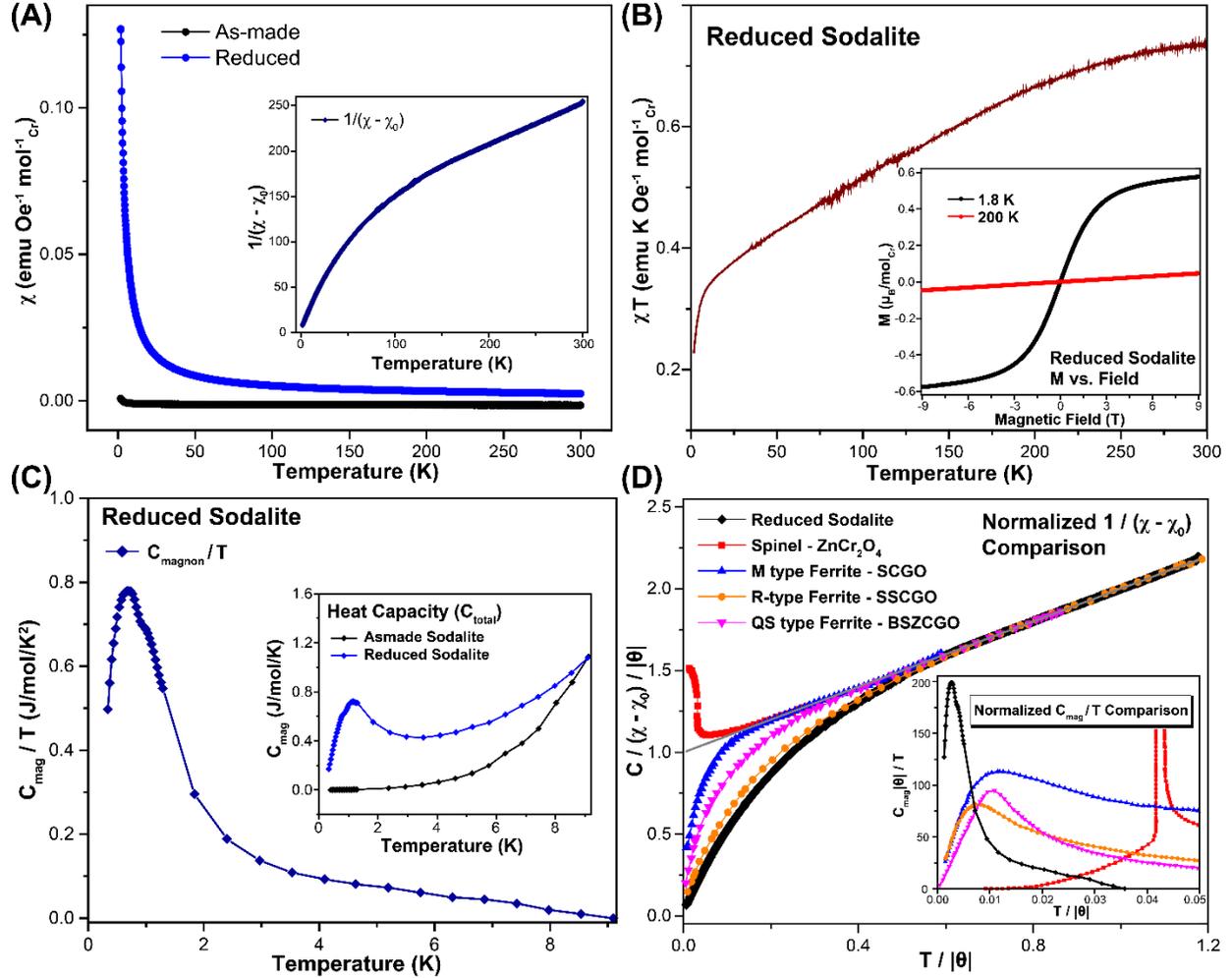

**Figure 3.** (A) Magnetic susceptibility characterization of the as-made and reduced Cr-sodalite zeolies, measured under an applied field of 5 kOe. $1/(\chi - \chi_0)$ is plotted versus $T$ for the reduced material in the inset; (B) $\chi T$ versus $T$ of the reduced sodalite zeolite. Inset: Field-dependent magnetization measurements for external field varied from -9 T to 9 T at 2 K and 200 K; (C) The lower temperature heat capacity, measured between 0.34 and 9 K under zero applied field of both sodalite samples (inset), and $C_{mag}/T$ of the reduced sodalite is plotted verses $T$ in the main panel with nonmagnetic contributions from the non-magnetic material subtracted from the total $C_p/T$, and integration was conducted to get $\Delta S_{magnetic}$, (D) Normalized inverse susceptibility (main panel) and normalized low-temperature heat capacity (inset) plotted for the geometrically frustrated chromates $ZnCr_2O_4$, the M-type ferrite SCGO, the R-type ferrite $SrSn_2Ga_{1.3}Cr_{2.7}O_{11}$ (SSCGO), the QS-type ferrite (BSZCGO), and the reduced CACr sodalite. The grey line in (A) shows the expected behavior for an ideal paramagnetic material with an antiferromagnetic intercept and no magnetic ordering, with a slope of 1 and a $y$-axis intercept of 1 [20]. Some of the data in these plots is taken from Ref. 19.

The Heat capacity ($C_p$) was measured down to 0.34 K on dense pellets of both the as-made Cr(VI) and the reduced Cr(III) sodalite materials. As shown in the Figure 3C inset, no change in entropy due to structural or magnetic transitions is observed for the as-made CACr sample, while two transitions (at around 0.8 K and 1.1 K respectively) are revealed in the heat capacity of the reduced CACr sodalite. It is not known whether the two transitions represent two different transitions in a single substance or whether there are slightly different magnetic ordering temperatures for the different ordering variants of the reduced zeolite phases observed by electron diffraction. $C_p/T$ of the reduced sodalite is plotted versus T in the main panel for further analysis. As for magnetic materials, the total heat capacity can be considered as a sum of magnetic contributions plus electronic, phononic, and other contributions.[24] To approximate the non-magnetic contributions, the heat capacity data from the non-magnetic sodalite is used as a reference and subtracted from $C_{total}$ to get $C_{magnon}$ of the reduced sodalite. By plotting and integrating $C_{magnon}/T$ in Figure 3C, it is found that the entropy change at the magnetic ordering transition ($\Delta S_{magnon}$) saturates at around 5 K with a total entropy change of about 1.5 J/mol/K, or about 25% of Rln2. This small value of $\Delta S_{magnon}$ indicates that either most of the entropy from magnetic ordering is not seen in this temperature range for reduced CACr, (in other words this compound does not show complete magnetic ordering down to 0.34 K.) or that there is some kind of diffuse loss of entropy occurring over a wide temperature range due to low dimensional ordering that is not easy to discern by this kind of measurement. What may be a magnetic ordering temperature of 1.1 K in the heat capacity data suggests a frustration index ($f = |\vartheta|/T_N$) of about 255/1.1 = 232, meaning that the spins don't freeze until they are supercooled to about 1/232$^{th}$ of the ordering temperature expected based on the Curie-Weiss theta. The system is then, at a minimum, very highly frustrated. If a macroscopic amount of entropy remains present down to the minimum measurement temperature of 0.34 K then the system is exotic indeed[21], and the alternative – a diffuse loss of entropy at higher temperatures without a dramatic signature in $\chi(T)$, although more "normal", would also be representative of unusual magnetic interactions.

An alternative, equally unusual explanation for the anomalous magnetic behavior observed could be that the Cr(III) present undergoes a spin state transition, i.e. due to a change from an orbital occupancy of one electron per $t_{2g}$ orbital at high temperatures, configuration $3d$ xy$^1$ xz$^1$ yz$^1$, to a lower unpaired spin $3d$ xy$^2$xz$^1$ configuration at low temperatures. Given that the electron configuration in Cr(III) in its insulating oxide compounds is very strongly governed by Hund's rule, this would be very surprising. Nonetheless to check for this possibility, XAS measurement was performed on the reduced green Cr(III) material.

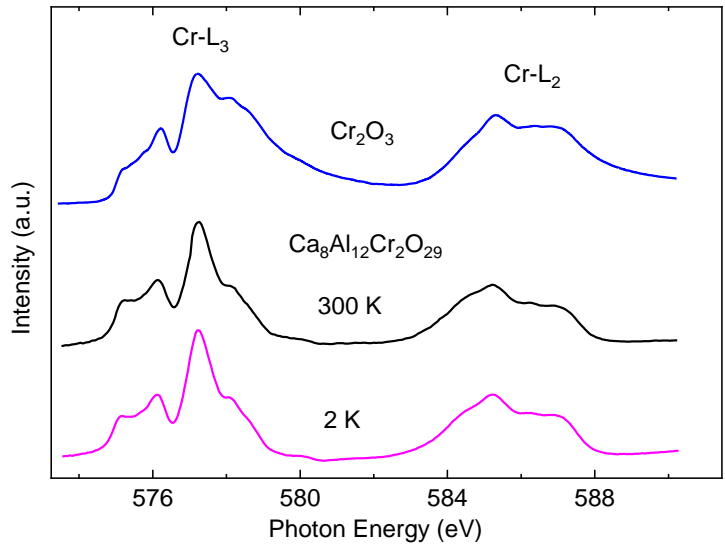

**Figure 4.** Experimental Cr-$L_{2,3}$ soft-XAS spectra of $Ca_8Al_{12}Cr_2O_{29}$ measured at 300 K and 2 K, and of $Cr_2O_3$ as $Cr^{3+}$ reference for comparison.

Figure 4 displays the XAS spectra at the Cr-$L_{2,3}$ edges of the reduced green material, $Ca_8Al_{12}Cr_2O_{29}$, together with that of $Cr_2O_3$. Soft XAS at the 3$d$ transition metal $L_{2,3}$ edges is known to be extremely sensitive to the valence[25], coordination[26], and spin-state[27,28] of the ion. The strong similarity of the spectra in Figure 4 therefore shows unambiguously that the Cr ions in the reduced zeolite $Ca_8Al_{12}Cr_2O_{29}$ are in the high-spin 3+ state and that the local coordination is octahedral. Nevertheless, the spectra of reduced sodalite $Ca_8Al_{12}Cr_2O_{29}$ and $Cr_2O_3$ are not identical. In fact, the $Ca_8Al_{12}Cr_2O_{29}$ spectra are more similar to that of Cr doped into $Al_2O_3$[29], which can be taken as an indication for rather short Cr-O distances and/or stronger distortions of the $CrO_6$ octahedra[29]. We cannot detect Cr with other valencies which would have not only very different spectral line shapes but also quite different energies[30,31], in our reduced zeolite. Of importance for the low temperature magnetic behavior of our reduced material is the observation that the 300 K and 2 K spectra of the $Ca_8Al_{12}Cr_2O_{29}$ are identical (also shown in Figure 4.). This proves that there is no spin-state transition on lowering the temperature since otherwise the spectra would have changed dramatically[27,28]. Thus, this confirmation of high-spin Cr(III) in the reduced sodalite sample by XAS measurement is in agreement with the $\mu_{eff}$ value calculated from the fitting to the high temperature range of the $1/(\chi - \chi_0)$ vs. $T$ data. By ruling out the possibility of a spin-state transition for the Cr, the strong, continuous deviation of $1/(\chi - \chi_0)$ vs. $T$ from Curie-Weiss behavior at low temperature can originate only from magnetic frustration.

In conclusion, zeolites are among the most important materials known and have been very widely studied. Here we report that the $Ca_8Al_{12}O_{24}(CrO_4)_2$ zeolite, when oxygen is removed to yield $Ca_8Al_{12}Cr_2O_{29}$, displays localized magnetic moment behavior, as is expected for the resulting $Cr^{3+}$-containing oxide. Analysis by electron microscopy reveals no significant impurity phase in the reduced material and that the elemental distribution is uniform in the reduced material. Soft x-ray absorption spectroscopy shows that the reduced material contains only $Cr^{3+}$ and that there is no change in the orbital occupancy and the spin-state of the Cr between room temperature and 2 K. The resulting electrically insulating zeolite has overall BCC symmetry, which is not generally considered to be a geometrically frustrating geometry, but the temperature-dependence of the magnetic susceptibility is exactly what is expected for a material that displays magnetic frustration, with the ratio of the Curie Weiss theta to the magnetic freezing temperature being approximately 232. We tentatively ascribe the magnetic frustration to either the interactions between disordered distorted $CrO_6$ octahedra in the cages or the overall geometric configuration of the resulting phase. The observation of magnetic frustration in a zeolite adds to the already remarkable set of properties that such materials can display.

**Experimental**

The CACr sodalite powder was synthesized using a traditional solid-state method. Stoichiometric $CaCO_3$, $Al_2O_3$, and $Cr_2O_3$ powders were mixed, ground, and annealed in air at 1100 °C for 72 h. Further grinding-and-annealing processes guarantee the purity. The reduced sodalite was obtained by annealing the as-made yellow sodalite powders under flowing forming gas (5% $H_2$ + 95% Ar) at 750 °C for two days. Carbothermal and borothermal reduction of the fully oxidized material was conducted by mixing carbon black or elemental boron powders with the yellow oxidized sodalite, pressing the mixtures into pellets and annealing at the same temperature for two days. Powder XRD patterns were collected using a Bruker D8 Advance Eco with Cu Kα radiation (λ= 1.5406 Å). Magnetization and heat capacity measurements were carried out using a Quantum Design PPMS (Dynacool), equipped with a vibrating sample magnetometer (VSM) option.

Samples for microstructure characterization were prepared by dispersing dry CACr sodalite particles on TEM grids (Lacey carbon films with no formvar). Selected area electron diffraction (SAED), conventional transmission electron microscope (TEM) imaging, high resolution TEM imaging and energy dispersive X-ray spectroscopic (EDS) mapping were performed on a double Cs-corrected FEI Titan Cubed Themis 300 scanning/transmission electron microscope (S/TEM), equipped with an X-FEG source and a super-X energy dispersive spectrometer (super-X EDS). The system was operated at 200 kV. The Cr-$L_{2,3}$ XAS spectra for $Ca_8Al_{12}Cr_2O_{29}$ and $Cr_2O_3$ were measured at the BL29 BOREAS beamline at the ALBA synchrotron radiation facility using the total electron yield mode.

**Acknowledgement**

The research on this material in the United States of America was supported by the US Department of Energy Division of Basic Energy Sciences, through the Institute for Quantum Matter, grant DE-SC0019331. This was the primary source of the funding for this work. Some of the synthetic work was performed in China during the COVID 19 pandemic, supported in part by the National Science Foundation of China (Grant 21805167). The authors acknowledge the use of Princeton's Imaging and Analysis Center, which is partially supported through the Princeton Center for Complex Materials (PCCM), a National Science Foundation (NSF)-MRSEC program (DMR-2011750). The work in Dresden was partially supported by the Deutsche Forschungsgemeinschaft through SFB 1143 (Project-ID 247310070).**References**

(1) T. Weller, M. Where Zeolites and Oxides Merge: Semi-Condensed Tetrahedral Frameworks. *J. Chem. Soc. Dalton Trans.* **2000**, *0* (23), 4227–4240. https://doi.org/10.1039/B003800H.

(2) A. Egerton, T.; Hagan, A.; S. Stone, F.; C. Vickerman, J. Magnetic Studies of Zeolites. Part 1.—The Magnetic Properties of CoY and CoA. *J. Chem. Soc. Faraday Trans. 1 Phys. Chem. Condens. Phases* **1972**, *68* (0), 723–735. https://doi.org/10.1039/F19726800723.

(3) A. Egerton, T.; C. Vickerman, J. Magnetic Studies of Zeolites. Part 2.—Magnetic Properties of NiA, NiX and NiY. *J. Chem. Soc. Faraday Trans. 1 Phys. Chem. Condens. Phases* **1973**, *69* (0), 39–49. https://doi.org/10.1039/F19736900039.

(4) Srdanov, V. I.; Stucky, G. D.; Lippmaa, E.; Engelhardt, G. Evidence for an Antiferromagnetic Transition in a Zeolite-Supported Cubic Lattice of F Centers. *Phys. Rev. Lett.* **1998**, *80* (11), 2449–2452. https://doi.org/10.1103/PhysRevLett.80.2449.

(5) Depmeier, W. The Sodalite Family – A Simple but Versatile Framework Structure. *Rev. Mineral. Geochem.* **2005**, *57* (1), 203–240. https://doi.org/10.2138/rmg.2005.57.7.

(6) Heinmaa, I.; Vija, S.; Lippmaa, E. NMR Study of Antiferromagnetic Black Sodalite $Na_8(AlSiO_4)_6$. *Chem. Phys. Lett.* **2000**, *327* (3), 131–136. https://doi.org/10.1016/S0009-2614(00)00859-9.

(7) Tou, H.; Maniwa, Y.; Mizoguchi, K.; Damjanovic, L.; Srdanov, V. I. NMR Studies on Antiferromagnetism in Alkali-Electro-Sodalite. *J. Magn. Magn. Mater.* **2001**, *226–230*, 1098–1100. https://doi.org/10.1016/S0304-8853(00)01286-5.

(8) Wang, Z.-X.; Shen, X.-F.; Wang, J.; Zhang, P.; Li, Y.-Z.; Nfor, E. N.; Song, Y.; Ohkoshi, S.; Hashimoto, K.; You, X.-Z. A Sodalite-like Framework Based on Octacyanomolybdate and Neodymium with Guest Methanol Molecules and Neodymium Octahydrate Ions. *Angew. Chem. Int. Ed.* **2006**, *45* (20), 3287–3291. https://doi.org/10.1002/anie.200600455.

(9) Lee, S.; Xu, H.; Xu, H.; Jacobs, R.; Morgan, D. Valleyite: A New Magnetic Mineral with the Sodalite-Type Structure. *Am. Mineral.* **2019**, *104* (9), 1238–1245. https://doi.org/10.2138/am-2019-6856.

(10) Hassan, I. Direct Observation of Phase Transitions in Aluminate Sodalite, $Ca_8[Al_{12}O_{24}](CrO_4)_2$. *Am. Mineral.* **1996**, *81* (11–12), 1375–1379. https://doi.org/10.2138/am-1996-11-1210.12